\documentclass[twocolumn,pre,amsmath,amssymb,floatfix]{revtex4}
\usepackage{epsf,graphicx}

\begin{document}
\title{Helicase on DNA: A Phase coexistence based mechanism}
\author{Somendra M. Bhattacharjee}
\affiliation{ Institute of Physics, Bhubaneswar 751 005 India\\ 
  somen@iopb.res.in }

\author{Flavio Seno}
\affiliation{ INFM-Dipartimento di Fisica, Universit\`a di Padova, Via
Marzolo 8, 35131 Padova, Italy\\
flavio.seno@pd.infn.it }
\date{\today }

\begin{abstract}
  
  We propose a phase coexistence based mechanism for activity of
  helicases, ubiquitous enzymes that unwind double stranded DNA.  The
  helicase-DNA complex constitutes a fixed-stretch ensemble that
  entails a coexistence of domains of zipped and unzipped phases of
  DNA, separated by a domain wall.  The motor action of the helicase
  leads to a change in the position of the fixed constraint thereby
  shifting the domain wall on dsDNA. We associate this off-equilibrium
  domain wall motion with the unzipping activity of helicase.  We show
  that this proposal gives a clear and consistent explanation of the
  main observed features of helicases.

\end{abstract}

\maketitle
\renewcommand{\thefootnote}{\fnsymbol{footnote}}

Nucleic acid helicases are defined as enzymes that translocate
directionally through double stranded nucleic acid substrates to
catalyze the separation of the complementary strands.  They facilitate
various biological processes such as DNA replication, recombination
and repair, RNA transcription, editing and splicing\cite{quattro}.
There are several structural varieties of helicases like monomeric
(e.g. PcrA), dimeric (e.g. Rep),  trimeric
(e.g. RecBCD), tetrameric (e.g. RNA polymerase) or closed hexameric
(e.g. DnaB), but all use the hydrolysis$^{\footnotemark[4]}$ of ATP to
ADP as the preferred source of energy
\cite{quattro,uno,due,tre,sspatel,porter98,soultanas2k,ha02}. 
\footnotetext[4]{In general, some nucleoside triphosphate (NTP) is
  required, e.g.  T7 gp4 can use GTP, SV40 large T antigen uses
  non-ATP nucleotides for unwinding RNA.\cite{comm1}}

 Bulk behaviour in solutions like average unwinding rates, step size,
average number of base-pairs opened per helicase, etc, are known for a
few helicases like Hepatitis-C virus helicase\cite{porter98}, PcrA
\cite{soultanas2k},  DnaB\cite{kim96,sspatel} and others.  Much attention
has recently been devoted\cite{dohoney,bianco01,bianco00} toward a
quantitative characterization of RecBCD enzymes, a multifunctional
trimeric protein complex (the products of the recB, recC and recD
genes\cite{Amnu}) that participates in the repair of chromosomal DNA
through homologous recombination.  In bacteria, like Escherichia coli,
RecBCD is involved, e.g., in protection against damages by UV or gamma
irradiation, and infection by bacteriophages.  In all cases, the full
functionality of RecBCD relies on the helicase and the nuclease
actions of its subunits.  The use of single-molecule,
micromanipulation tools allowed for monitoring in detail
translocation\cite{dohoney,bianco00}, unwinding \cite{bianco01} and
processivity (rate of dissociation) \cite{bianco01} of individual
RecBCD enzyme molecules on dsDNA.  Such experiments elucidated several
new aspects of helicase behaviour and showed that many properties
could be more related to general principles than on specific chemical
details.  In particular it has been observed that (a) RecBCD unwinds
dsDNA at a uniform rate, over a wide range of ATP concentrations, as
it moves on one strand, (b) the nuclease activity does not affect
unzipping, and c) the helicase can work in presence of DNA gaps upto
certain lengths.   More recently, winding-rewinding for E.Coli Rep
helicase-DNA complex\cite{ha02} has been observed at a single molecule
level.  

Despite these varieties of experimental findings no clear mechanism
coupling the motor action and the helicase activity is known yet. To
fill this gap, in this paper we present a simple, but powerful
argument, based on the principle of phase
coexistence\cite{smb,maren_dyn,maren_pre,smb_maren,otherref}, that
provides clear and robust explanations to the gross observed features.
There are a few biological operational models built on how a helicase
presumably might work\cite{sspatel}.  The analysis reported here gives
a thermodynamic basis to a model called the ``wedge model'' according
to which the motion of the helicase ``provides enough force to enable
the helicase to destabilize the base pairs at the junction by a
process resembling the action of a wedge''\cite{sspatel}.  In our
proposed mechanism, energy (from ATP) is required for translocation
activity or the motor action of the helicase and not directly for base
pair breaking and therefore, according to the classification scheme of
Ref. \cite{due}, this corresponds to passive helicases.  Additional
features required for active helicases are ignored in this first
study. Our proposal is supported by computer simulation of an exactly
tractable model.  To our knowledge, this is the first theoretical
study of the dynamics of a DNA-helicase complex.

\begin{figure}[tbp]
\centerline{\includegraphics[clip,width=3.25in]{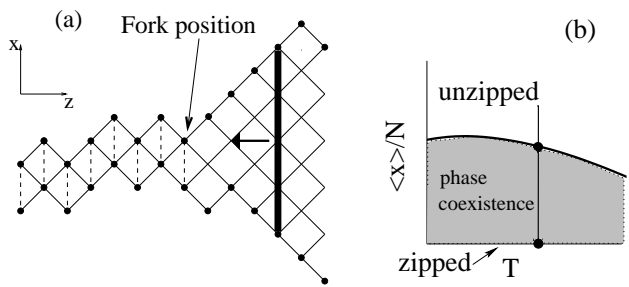}}   
\caption{(a) A typical configuration of the simulated DNA 
  with the helicase (thick rod) as modeled on a square lattice (thin
  lines). Bases are represented by dots and paired bases are shown by
 dotted lines. The position of the fork coincides with the last
  paired base.  The arrow is indicating the motion of the helicase.
  Except for the rigid hard-core constraint, there is not other direct
  interaction between the DNA and the helicase.  (b) Schematic phase
  diagram in the separation (stretch) vs temperature plane.  At fixed
  temperature, a finite end-point separation leads to a coexistence of
  the two phases indicated by the filled circles. The interface of the
  two phases is the domain wall.}
\label{Fig1}
\end{figure}

We study the joint dynamics of the helicase and the DNA in a two
dimensional fork model (Y-Model) \cite{maren_dyn,maren_pre}.  The two
strands of DNA are represented by two directed and mutually avoiding
walks formed by $N$ bases each.  In two dimensions, on the square
lattice (see Fig. \ref{Fig1}) this implies that the two walks follow
the positive direction of the diagonal axis ($z$); in other words the
coordinate along such direction always increases.  The perpendicular
direction $x$ measures, in unit of the elementary square diagonal, the
distance between the two complementary monomers belonging to the two
strands.  When this distance is equal to $1$ they are considered in
contact: a binding energy is gained which is uniform ($\epsilon =1$)
for a homogeneous model of DNA (homo-DNA) but chosen randomly from two
different values ($\epsilon_1, \epsilon_2$) for a heterogeneous DNA
(hetero-DNA).  Notice that due to the geometrical properties of the
lattice the two complementary monomers are labeled by the same
$z$-coordinate, as one would require for base pairing in DNA.  In the
Y-model unzipping can occur only processively, e.g.  bubbles are
suppressed along the chain: the only conformations considered have the
first $N-m$ monomers bounded, whereas the remaining $m$ are separated
in a $Y$-like conformation.  The fact that the Y-model does not allow
rejoining of the unzipped portion of the dsDNA is similar to the
geometry observed in the experiment of Ref \cite{bianco01} (see also
Fig. 4 of Ref.  \cite{Henn}).  Also, bubbles are suppressed for DNA at
temperatures much below its melting temperature $T_m$, temperature
below which the two strands are zipped.  In the case of homogeneous
interaction the exact phase diagram and other static and dynamical
quantities can be exactly determined (also in the presence of a
stretching force)\cite{maren_dyn,maren_pre}.

The coarse-grained nature of the model\cite{PS} needs to be stressed
here.  Monomers are to be thought of as groups of bases, and ignoring
helicity or restricting to two dimensions are more for simplification
of the calculation than artifacts$^{\footnotemark[4]}$
\footnotetext[4]{The qualitative features of
the unzipping phase transition of Refs.
\cite{smb,maren_dyn,maren_pre} are observed in more complex models
in \cite{enzo}}.  Such coarse-grained models, or even simpler ones,
are used in various DNA related problems\cite{otherref,causo} and even
in analysis of thermal melting of DNA\cite{daune}. The spirit behind our approach is that the key
element that can influence universal behaviour of helicase
translocation is the competition between a Y-fork conformation which
can be unzipped by paying energy and the movement of an opening
machine.

Several studies of theoretical DNA models
\cite{smb,maren_dyn,maren_pre,smb_maren,enzo,otherref} have
established the existence of a sharp unzipping phase transition of a
dsDNA at a critical stretching force applied at one end on the two
strands. This implies that in the conjugate ensemble of a
fixed-separation constraint for the two strands, there is a
coexistence of domains of zipped and unzipped phases.  For the
fixed-force ensemble for a homo-DNA in the large length limit
($N\rightarrow\infty$), the critical force $g_c$ for the
zipping-unzipping transition, and the end separation under a force
$g>g_c$ are given by\cite{maren_pre,maren_dyn}
\begin{equation}
\label{eq:xg}
g_c(T)= \frac{T}{2} \  \cosh^{-1} (e^{1/T} -1), {\rm and}\quad
\frac{\langle   x\rangle}{N} = \tanh \frac{g}{2T},  
\end{equation}
\noindent where temperature $T$ is measured in units of
$k_B/\epsilon$, $k_B$ being the Boltzmann constant.  From these
equations the phase coexistence curve $\langle x\rangle/N$-vs-$T$ can
be determined exactly and it is schematically shown in Fig.
\ref{Fig1}b. Under a fixed-distance constraint, represented by the
vertical line in Fig \ref{Fig1}b, the DNA chain splits into domains of
zipped and unzipped phases.  The length of the unzipped strand can be
read off from the upper line of the coexistence curve.  This fact can
be checked independently and we have verified it not only for the
Y-model but also for models that allow bubbles.

\begin{figure}[tbp]
\includegraphics[width=2.5in,clip]{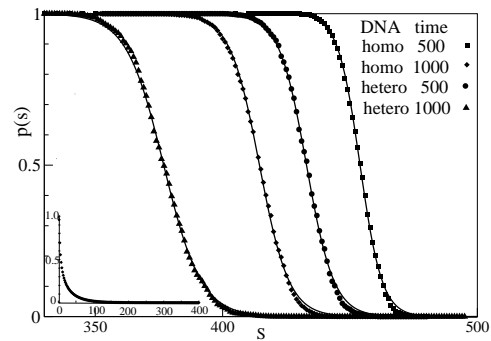}
\caption{
  Zipping probabilities at two different times (after thermal
  averaging) for the homo-DNA ($\epsilon=1$) and hetero-DNA
 ($\epsilon_1=1,\epsilon_2=0.5$).  The solid lines are fits to Eq.
  \ref{eq:1}.  The inset shows the equilibrium zipping probablities (no
  domain wall) for
  the case of $\epsilon=0$ when the ends at $z=400$ are kept separated. }
\label{Fig3}
\end{figure}

We now make our hypothesis.  If we think that the helicase, by virtue
of its size larger than the the separation of the two DNA strands and
the excluded volume interaction, acts as a geometrical separator, {\it
  the helicase-DNA complex constitutes exactly a fixed-stretch
  ensemble}.  The unwinding activity can then be simply associated
with the motion of the domain wall which necessarily forms and follows
the motor action of the helicase.  In other words, the helicase plays
a double role: firstly, its presence thermodynamically implies the
existence of a domain wall, and secondly, its translocation induces a
motion of the domain wall to reach its equilibrium position.  Such an
approach puts the primary role on the translocation motion.  The
thermodynamic force, that drives the domain wall toward its
equilibrium position, provides the mechanism for base pair opening.

To verify our hypothesis we have numerically studied the DNA model
described above mimicking the effects of the helicase with a rod of
length $l$ (see Fig. \ref{Fig1}a).  Dynamics is introduced by a
Monte-Carlo procedure.  For the DNA, one among the $2N$ monomers is
randomly chosen and an attempt is made to modify its position with
respect to all the others (which remain fixed).  The move can in
principle increase or decrease by one unit the distance between the
strands. The move is accepted according to standard Metropolis rules.
A Monte-Carlo unit time is defined as $2N$ single-monomer attempted
moves.  The scaling properties of this DNA dynamics, also in the
presence of a stretching force, have already been
determined\cite{maren_dyn}.  The helicase moves forward (motor action)
by unit step, along the $-z$ direction, on the DNA, if it is not
hindered by the chain configuration (excluded volume interaction).
The motion is kept unidirectional as found for RecBCD in Ref.
\cite{dohoney} (see below also).  The motion of the helicase is
attempted at every Monte Carlo step.  In our simulation we start with
a configuration where the helicase is attached to one end of a dsDNA (
as in experiments of Ref \cite{dohoney,bianco01,bianco00}).  The
temperature is maintained constant for the DNA to be in the zipped
phase (below $T_m$).  The overall dynamics is off-equilibrium.  Notice
that there is no specific interaction but both DNA and helicase
dynamics are constrained by the excluded volume.

Throughout the simulation we  monitored the position of
the last zipped base pair (fork position see Fig. \ref{Fig1}a), the
zipping probability $p(s,t)$ that a base pair at site $s$ at time $t$
is zipped, and the average position of the helicase at time $t$. 
In most simulations the length $l$ of the helicase was taken equal to
6 (though we studied up to $l=14$) whereas 1000 thermal averages were
necessary. The length $N$ of the strands were varied according to
the different experiments ranging up to $N=1000$.
For an analysis of the domain wall, the zipping probability $p(s,t)$
can be fitted by a function: 
\begin{equation}
\label{eq:1}
p(s,t)=({1}/{2})\ [1 - \tanh\{{\bf (}s-s_0(t){\bf )}/{w(t)}\}   ]
\end{equation}
\noindent where $s_0(t)$ and $w(t)$ are the position and the width,
respectively, of the domain wall.

\begin{figure}[tbp]
\centerline{\includegraphics[width=3.in,clip]{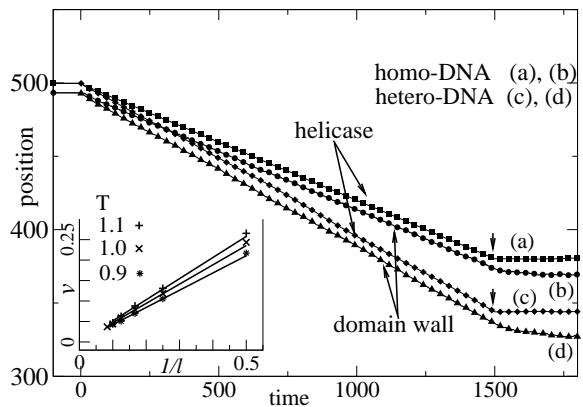}}   
\caption{ Positions of the helicase and  the domain wall  as
  a function of time for  (a,b) homo- and (c,d) hetero- DNA. The size
  of the helicase is $l=6$.  When the helicase is stopped
  at time= 1000 (indicated by the arrows),   the domain
  wall evolves toward the equilibrium position. The inset shows  the
  speed of the helicase for  different temperatures  and lengths, with
  the solid lines representing $v=1.28 [\exp(-1/T)]/l$.} 
\label{Fig4}
\end{figure}

In Fig \ref{Fig3} we plot for two different times the zipping
probability $p(s,t)$ for the homogeneous and for an heterogeneous
case. In all cases the domain wall behaviour predicted is well fitted
by a tanh profile (Eq. \ref{eq:1}).  We point out that such domain
walls are not found if the DNA strands are noninteracting, {\it i.~e.}
$\epsilon=0$ (see inset of Fig. \ref{Fig3}).  In Fig. \ref{Fig4} we
show the instantaneous positions 
of the helicase and the domain wall (computed through Eq.  \ref{eq:1})
as a function of time. Movement starts at $t=0$ when the helicase is
loaded. The two quantities proceed uniformly and cooperatively through
the DNA unwinding it.  
 The domain wall evolves toward the equilibrium position whenever the
helicase motion is stopped and this position
turns out to be not very far away from the instantaneous position.
This indicates an adiabatic adjustment of the domain wall to the
instantaneous position of the mobile helicase.  We stress that had
there been no interaction between the DNA and the helicase, the
latter, as a phantom motor, would have moved with the assigned speed
with the DNA remaining bound (since we are below $T_m$).  It is
important to compare this motion  with the unzipping dynamics in a fixed force
ensemble which shows a characteristic scaling
behavior\cite{maren_dyn}, namely a nonlinear evolution $m(t) \sim
t^{1/3}$, where $m(t)$ is the number of unzipped bases at time $t$.
In contrast, and this is the central point of our work, we find that
the combined dynamics involving the excluded volume interaction
between the helicase and the DNA (but no external force) leads to the
uniform motion of both the helicase and the domain wall.  Our results
of Fig.  \ref{Fig4} should be compared with Fig. 2 of Ref
\cite{bianco01}.  The effective velocity is smaller than the
unhindered one and depends on the size $l$ of the helicase, the
temperature and the sequence. The $l$ and $T$ dependence can be
estimated if we think that to move the helicase we have to unzip a
base pair at the fork position, and to allow the formed kink to reach
the position of the helicase.  The base pair is broken with a
probability $\exp({-\frac{1}{T}})$ and the kink needs a time of order
$l$ to reach the helicase.  Therefore the velocity is proportional to
$[{\exp{(-\frac{1}{T})}}]/{l}$ (confirmed by the fit of our data, see
the inset of Fig \ref{Fig4}.). A strong dependence on $T$ has also
been found in Ref. \cite{bianco01}.  In our simulation no sequence
dependent nonuniformity in motion was ever discernible so long as the
heterogeneity was uncorrelated.  The velocity we observed, originating
solely from the motor action, is a lower bound because any periodic
conformational change of the helicase\cite{dohoney} during its motion
(ignored here mainly to illustrate the role of the domain wall) would
assist the motion of the domain wall itself.

Our proposal lends itself to several predictions. We discuss a few
here.  Since the position $s_0(t)$ of the wall is determined by the
location of the stretching constraint put up by the helicase, the
unzipped part beyond the helicase should not affect the action of the
helicase. We verified this explicitly in our simulation: a part of the
unzipped DNA beyond the helicase is chopped off (which mimics RecD
activity\cite{dohoney}) at arbitrarily chosen times and there is
neither any change in the nature of the wall (e.g. $p(s,t)$) nor in
the rate of unzipping. This agrees with the observations in Refs.
\cite{dohoney,bianco01} that the nuclease activity does not modify the
unwinding action of RecBCD.  Next, the width of the wall,  as
  defined by Eq. \ref{eq:1},  gives a
length-scale for the helicase activity.  The sequence randomness in a
heterogeneous chain does not matter if there is no correlation beyond
this scale, as found both in simulation and real experiments.  We have
seen a periodic modulation (not shown) in the helicase motion if the
DNA sequence is periodic of two pairing energies with periodicity
larger than the width of the domain wall.  Another consequence of this
scale is that a nick or break smaller than the domain wall width will
not be recognized, providing an interpretation of the observations in
Ref \cite{bianco00} regarding the size of the gap a helicase can
negotiate.

\begin{figure}[tbp]
\centerline{\includegraphics[width=2.5in,clip]{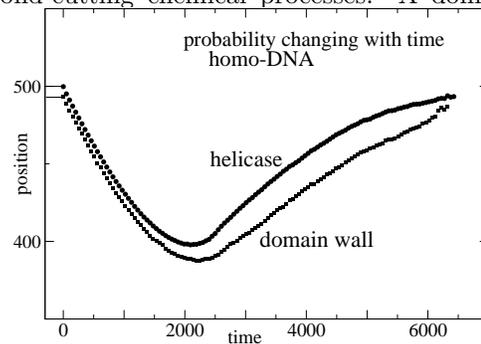}}   
\caption{ Positions helicase ($l=6)$ for an homogeneous DNA
with a biased probability $P(t)$. Probability decreases linearly
from $1$ for $t=0$ to $\frac{1}{2}$ for $t\approx 2500$ after which
it remains constant at  $1/2$}
\label{Fig5}
\end{figure}

We have also simulated cases where the helicase undergoes a biased
random walk-type motion as expected for RecQ\cite{harmon}. We
introduced a probability $P(t)$ that the helicase could step along the
$-z$ direction on the DNA (when this is sterically acceptable) but
also a probability $1-P(t)$ that it could move in the opposite
direction ({\it i.~e.}, away from the domain wall).  When $P(t)$ is
kept fixed at a value $P_0$ in the interval $ 1 \geq P_0 \geq
\frac{1}{2}$ ($P_0=1$ being the case studied in the first part of the
paper) we find again a uniform motion with velocity related to $P_0$.
At $P_0 \equiv \frac{1}{2}$ (random walk) situation changes: unzipping
does not proceed and the helicase dissociates from the double strand.
This behaviour is illustrated in \ref{Fig5} where we are plotting the
position of the helicase as a function of time: $P(t)$ decreases
linearly from $1$ down to $\frac{1}{2}$ and then remains constant.  As
expected the unwinding proceeds non-uniformly until the random walk
regime is reached. At that time the helicase is discarded and the DNA
zips again.  It suggests,  though a bit speculative,  that a
probability affecting the forward 
motion of the helicase could be a phenomenological characterization of
the gradual wobbliness of the helicase on the track. 

In conclusion, we have shown that associating the helicase activity
with the domain wall motion in a fixed-stretch ensemble accounts for
several observed features, as e.g., the uniformity of unzipping, no
sequence-dependent nonuniformity, and the insensitivity to the
nuclease action, without any requirement of extra specific
bond-cutting chemical processes.  A domain wall also gives a
quantitative meaning to the Y-fork in the terminology of DNA
replication.  The underlying thermodynamic basis gives a robustness to
the mechanism that it could be at work for hexameric helicases also.


\noindent FS was supported by MIUR-COFIN01. FS thanks kind
hospitality of Institute of Physics, Bhubaneswar. 


\end{document}